 \documentclass[10pt,preprint]{aastex}  
 \usepackage{natbib}
\def\ang{\AA}

\def\gapprox{\lower.4ex\hbox{$\;\buildrel >\over{\scriptstyle\sim}\;$}}
\def\lapprox{\lower.4ex\hbox{$\;\buildrel <\over{\scriptstyle\sim}\;$}}
\def\ref#1{\par\noindent\hangindent1cm {#1}}

\shortauthors{ASCHWANDEN}
\shorttitle{Global Energetics of Solar Flares}

\begin{document}

\title{ Global Energetics of Solar Flares: X. Petschek Reconnection Rate 
	and Alfv\'en Mach Number of Magnetic Reconnection Outflows }

\author{Markus J. Aschwanden}

\affil{	Solar and Astrophysics Laboratory,
	Lockheed Martin Advanced Technology Center, 
        Dept. ADBS, Bldg.252, 3251 Hanover St., Palo Alto, CA 94304, USA; 
        (e-mail: \url{aschwanden@lmsal.com})}

\begin{abstract}
We investigate physical scaling laws for magnetic energy dissipation
in solar flares, in the framework of the Sweet-Parker model
and the Petschek model. We find that the total dissipated magnetic energy
$E_{diss}$ in a flare depends on the mean magnetic field component
$B_f$ associated with the free energy $E_f$, the length scale $L$ of the
magnetic area, the hydrostatic density scale height $\lambda$ of
the solar corona, the Alfv\'en Mach number $M_A=v_1/v_A$ (the ratio 
of the inflow speed $v_1$ to the Alfv\'enic outflow speed $v_A$), and the
flare duration $\tau_f$, i.e., $E_{diss} = (1/4\pi) B_f^2\ L\ \lambda\ 
v_A\ M_A\ \tau_f$, where the Alfv\'en speed depends on the nonpotential
field strength $B_{np}$ and the mean electron density $n_e$ in the
reconnection outflow. Using MDI/SDO and AIA/SDO observations and 
3-D magnetic field solutions obtained with the vertical-current 
approximation nonlinear force-free field code (VCA-NLFFF) we measure
all physical parameters necessary to test scaling laws, which
represents a new method to measure Alfv\'en Mach numbers $M_A$,
the reconnection rate, and the total free energy dissipated 
in solar flares.

\end{abstract}


\section{	INTRODUCTION 			}

Key observations that provide evidence for magnetic reconnection in
solar flares have been furnished mostly from Yohkoh observations: 
(i) the X-point geometry that is visible as cusp
in long-duration events (Tsuneta et al.~1992);
(ii) the altitude of the X-point as observed in above-the-loop-top
hard X-ray sources (Masuda et al.~1994) and with time-of-flight
measurements (Aschwanden et al.~1996);
(iii) the rise of the X-point with time that maps to separating
footpoints and double ribbons (Sakao et al.~1998; Fletcher and Hudson 2001);
(iv) the horizontal symmetry as observed in simultaneous hard X-ray 
emissions from conjugate footpoints (Sakao 1994);
(v) the vertical symmetry as observed in bi-directional type III bursts
(Aschwanden et al.~1995), in correlations between hard X-ray pulses 
and type III bursts (Aschwanden et al.~1993),
and in vertically symmetric hard X-ray sources below and above 
the X-point (Sui and Holman 2003);
(vi) post-reconnection relaxation measured from the loop shrinking
ratio (Forbes and Acton 1996) and the progression of cooling loops
below hot loops (Svestka et al.~1987);
(vii) reconnection inflows detected as inward motion in EUV emission
(Yokoyama et al.~2001);
(viii) reconnection outflows observed as supra-arcade downflows
(McKenzie and Hudson 1999);
(ix) slow-mode standing shocks in high-temperature ridges of the
flare arcade (Tsuneta 1996);
and (x) fast-mode standing shocks indicated by high densities above
looptops (Tsuneta 1997b) and by above-the-loop-top hard X-ray
sources (Masuda et al.~1994). 
Most of these observations are consistent with the standard flare
scenario of Carmichael (1964), Sturrock (1966), Hirayama (1974),
Kopp and Pneuman (1976), Tsuneta (1996; 1997a; 1997b), and Shibata et al.~(1995),
also known as CSHKP model (named after the initials of the first five
authors).

While most of the previously listed key observations capture the
ongoing physics in solar flare magnetic reconnection processes in
a qualitative way, based on 2-D images in soft X-rays and EUV, 
more quantitative models can be established by physical scaling laws.
A first step is the evaluation of which physical parameters are relevant
for the derivation of a scaling law, and the second step involves
fitting theoretical scaling laws to observations. 
A compilation of scaling laws for coronal heating processes has been
pioneered in Mandrini et al.~(2000), which are also applicable to solar 
flare heating processes. Their study includes a variety of 22
theoretical scaling laws of the power law form 
$\varepsilon \propto B^a L^b \rho^c v^d R_c$, with 
$\varepsilon$ the volumetric heating rate, $B$ the average
coronal field strength, $L$ a length scale of loops, $\rho$ the
mass density, $v$ the transverse velocity at the chromospheric footpoints,
and $R_c$ a characteristic horizontal length scale for the magnetic
field or flow. The application of these scaling laws to observations
in Mandrini et al.~(2000), however, has some shortcomings: 
(i) the mean magnetic field $B$
is parameterized as a function of the loop length $L$, i.e., $B(L) 
\propto L^{\delta}$, but the power law slope $\delta$ differs for small loops
(in the near-field zone) and large loops (in the far-field zone), and
thus invalidates their power law "ansatz"; (ii) The magnetic field component
$B$ is measured from potential and non-potential force-free field
extrapolations, but ignores the fact that only the free (magnetic)
energy can be dissipated  into other forms of energy in coronal heating
or flare events; and (iii) other physical parameters that turn out to be
important for a realistic scaling law are ignored, such as the 
magnetic field associated with the free energy ($B_f \propto \sqrt{E_f}$),
the Alfv\'en speed in the reconnection outflows ($v_A$), the reconnection 
inflow speed ($v_1 = v_A M_A$), the Alfv\'en Mach number ($M_A$), or 
the hydrostatic electron density scale height ($\lambda$).
Although the starting point of the study of Mandrini et al. (2000) deals 
with non-flaring loop heating, rather than solar flare processes that are the 
focus of this study, both phenomena are produced by magnetic reconnection 
processes, and thus can be quantified with the same scaling laws. We will address
more comprehensive scaling laws for magnetic reconnection in solar flares
in this study that improve on these shortcomings.

What is the progress in modeling physical scaling laws of magnetic
reconnection processes in solar flares, compared with the observational
highlights during the Yohkoh mission? Modeling of the 3-D geometry of
the space-filling magnetic field ${\bf B}=[B_x(x,y,z), B_y(x,y,z),
B_z(x,y,z)$], which beyond the potential field, calculates force-free
field solutions of the nonpotential field and the free energy, 
containing the only amount that can be dissipated in solar flares,
have become available in the recent decade only. In this study,
based on the data analysis from Paper I and Paper IX, magnetic parameters
have been computed with the {\sl Vertical-Current Approximation
Non-Linear Force-Free Field (VCA-NLFFF)} code, which determines
the free energy from the current-driven helical twisting of
magnetic field lines. Typically, the azimuthally twisted magnetic
field component ${\bf B}_f = {\bf B}_\varphi$ is a factor of
$B_f/B_{np} \approx 0.3\pm0.1$ smaller than the co-spatial 
nonpotential field strength, which implies that the free
energy is by a factor of $E_f/E_{np} \approx 0.10\pm0.07$ smaller.
Thus the discrimination of nonpotential and free energy is important
in the energy budget of solar flares. On the other hand, the Alfv\'en
velocity $v_A$ has to be computed based on the (total) nonpotential 
magnetic field strength $B_{np}$, rather than on the azimuthally
twisted field component $B_f$ that is associated with the free energy.
 
In this study we address mostly the Sweet-Parker model and the
Petschek model applied to solar flares, but there are also 
applications of Petschek-type reconnection to the magnetopause 
and solar wind,
dealing with unsteady Petschek reconnection (Biernat et al.~1987), 
resistivity enhancement by instabilities (Kulsrud 2001),
current-driven anomalous resistivity (Uzdensky 2003; B\"uchner 2006),
or Petschek-type reconnection exhausts in the solar wind
(Gosling et al.~2006a, 2006b).

The content of this paper includes a theoretical description of
the Sweet-Parker current sheet model and the Petschek model
(Section 2), observations and data analysis of M- and X-class
solar flare events (Section 3), a discussion of observational
evidence for reconnection inflows, the Alfv\'en Mach number,
flare ribbon motions, and scaling law parameterization
(Section 4), and conclusions (Section 5).

\section{	THEORY OF SCALING LAWS 	}

In the following sections we define theoretical 
scaling laws that quantify the volumetric energy dissipation rate 
$\varepsilon_{diss}$ (in units of [erg cm$^{-3}$ s$^{-1}$])
as a function of physical parameters measured during flare events, 
such as the coronal mean nonpotential  magnetic field strength $B_{np}$,
the magnetic field component $B_f$ associated with the free energy, 
the length scale $L$ of the magnetic field (fractal) area, 
and the electron density $n_e$.  
The free energy component is generally defined by the 
difference between the nonpotential and potential field component,
which can be represented by a vector difference equation,
${\bf B_f={\bf B}_{np} - {\bf B}_p}$, or by an energy difference
equation, $B_f^2=B_{np}^2 - B_p^2$, and quantifies the
maximum energy that can be dissipated during a flare. 
We define the volumetric energy dissipation rate $\varepsilon_{diss}$ by
\begin{equation}
	\varepsilon_{diss} \propto L^a\ B_{np}^b\ B_f^c\ n_e^d 
	\qquad {\rm [erg\ cm}^{-3}\ {\rm s}^{-1}] \ ,
\end{equation}
in the form of power law dependencies (with power exponents 
$a, b, c, d$). Since we are interested in the total energy 
dissipated in a flare event, we have to integrate 
the volumetric dissipation rate $\varepsilon_{diss}$ 
(in units of [erg cm$^{-3}$ s$^{-1}$])
over the flare volume $V$ and flare duration $\tau_f$,
\begin{equation}
	E_{diss} = 
	\varepsilon_{diss}\  V\ \tau_f \qquad {\rm [erg]} \ .
\end{equation}
We characterize the scaling of the flare volume $V$ as a function 
of the length scale $L$ with the Euclidean 3-D dimension,
\begin{equation}
	V(L) = L^3 \qquad {\rm [cm}^3] \ .
\end{equation}
With these definitions (Eqs.~1-3), the total dissipated energy $E_{diss}$ 
can be expressed as
\begin{equation}
	E_{diss} \propto L^{(a+3)}\ B_{np}^b\ B_f^c\ n_e^d\ \tau_f 
	\qquad {\rm [erg]} \ .
\end{equation}
Note that the scaling laws are different for the two energy parameters
$\varepsilon_{diss}$ and $E_{diss}$, due to the length dependence
of the flare volume $V(L)$. For completeness we define also the
Poynting flux $P$, which is defined by the energy input rate
or dissipation rate per unit area,
\begin{equation}
	P_{diss} = \varepsilon_{diss}\ L 
	\propto L^{(a+1)}\ B_{np}^b\ B_f^c\ n_e^d 
	\qquad {\rm [erg\ cm}^{-2}\ {\rm s}^{-1}] \ .
\end{equation}

\subsection{	The Sweet-Parker Current Sheet  	}

One of the first theoretical concepts of a magnetic reconnection
process occurring in a solar flare is the so-called Sweet-Parker
current sheet (Sweet 1958; Parker 1963, 1988). The theoretical
derivation of this process can be found in recent textbooks
(e.g., p.120 in Priest and Forbes 2000; p.409 in Aschwanden 2004).
A main assumption is that the diffusion region is much longer
than wide, $\Delta \gg \delta$ (Fig.~1, left). For steady, 
compressible flows ($\nabla \cdot {\bf v} \neq 0$), it was found 
that the reconnection outflow $v_2$ roughly has Alfv\'en speed,
\begin{equation}
        v_2 \approx v_A = {B_2 \over \sqrt{4 \pi m_p n_e}} \ ,
\end{equation}
Observationally, the outflow speed $v_2$ is slower than the 
expected Alfv\'en velocity $v_A$. Since the Alfv'en speed scales 
reciprocally with the square root of the electron density, $v_A 
\propto \sqrt{n_e}$, a lower value of $v_A$ could be explained by 
larger electron densities $n_E$ in flaring reconnection regions.
The outflow speed $v_2$ relates to the inflow speed 
$v_1$ reciprocally to the cross sections $\delta$ and $\Delta$ 
and mass densities $\rho_1=m_p n_1$ and $\rho_2=m_p n_2$
(according to the continuity equation),
\begin{equation}
        {\rho}_1 {v}_1 \Delta =  {\rho}_2 {v}_2 \delta \ ,
\end{equation}
and that the {\sl reconnection rate} $M_A$, defined as the Mach 
number ratio of the external inflow speed $v_1$ to the (Alfv\'en) 
outflow speed $v_A$, is 
\begin{equation}
        M_A = {v_1 \over v_A} = {1 \over \sqrt{S}} \ .
\end{equation}
The {\sl Lundquist number} $S$ (or {\sl magnetic Reynolds number $R_m$}) 
is defined by
\begin{equation}
        S={ {v}_A L / \eta} \ ,
\end{equation}
analogous to the Reynolds number $R = v L / \eta$ defined for a general
fluid velocity $v$. 

So, for typical coronal conditions (with a large Lundquist number of 
$S = R_m \approx 10^8-10^{12}$) the reconnection rate is typically 
$M_A \approx 10^{-4}-10^{-6}$, which yields inflow speeds in the 
order of $v_1 \approx v_A M_A \approx 1000$ km s$^{-1}$
$\times 10^{-5} \approx 0.1$ km s$^{-1}$ and predicts 
extremely thin current sheets with a thickness of $\delta = 
\Delta (v_1/v_A) \approx \Delta \times 10^{-5}$.
So, a current sheet with a length of 
$\Delta \approx 10,000$ km would have a thickness of only 
$\delta \approx 0.1$ km. The predicted total energy $E_{diss}$ dissipated
during a flare with duration $\tau_f$ is (with the label MR1
denoting the first type of magnetic reconnection model), 
\begin{equation}
	E_{MR1} = \int \left( {dE \over dt} \right) \ dt
	\approx 2\ {B_f^2 \over 8\pi} {dV \over dt} \ \tau_f
        \approx {B_f^2 \over 4\pi}  L^2\ v_1     \ \tau_f
        \approx {1 \over 4\pi} B_f^2\ L^2\ v_A\ M_A\ \tau_f 
	\ . 
\end{equation}
The term $dV/dt=L^2 v_1$ decribes the inflow rate of the plasma volume
into the reconnecting diffusion region (Fig.~1 top). The factor 2 accounts 
for the symmetric plasma inflows on both sides of a vertical current sheet.
The volume is normalized to an Euclidean 3-D cube with length scale $L$ and
filling factor of unity, i.e., $V = L^3$ (Eq.~3).

The mean volumetric heating rate $\varepsilon$ is 
defined by dividing the total dissipated energy by the flare
volume $V$ and by the flare duration $\tau_f$ (Eq.~2), 
which yields a scaling law of
\begin{equation}
	\varepsilon_{MR1} 
	= {E_{MR1} \over V \tau_f}  
	= {E_{MR1} \over L^3\ \tau_f}  
        = {1 \over 4\pi} B_f^2\ L^{-1}\ v_A\ M_A \ .
\end{equation}
For typical solar flares parameters, which we take from the median
values of the parameter set analyzed in the following Section and
listed in Table 1, ($B_{np}=240$ G, $B_f=70$ G, $L=3.4 \times 10^9$ cm, 
$n_e=9 \times 10^{10}$ cm$^{-3}$, $\tau_f=900$ s, $S=10^{10}$, 
we obtain (with Eq.~10) typical values of 
$M_A=1/\sqrt{S} = 10^{-5}$ for the Alfv\'en Mach number, 
$v_A=1700$ km s$^{-1}$ for the Alfv\'en speed, 
$E_{diss} \approx 7 \times 10^{27}$ erg 
for the total dissipated energy, and (with Eq.~11) a
volumetric energy dissipation rate of 
$\varepsilon_{diss} \approx 2 \times 10^{-4}$ erg cm$^{-3}$ s$^{-1}$,
which corresponds to a Poynting flux of $P = \varepsilon_{diss} \ L
\approx 7 \times 10^{5}$ erg cm$^{-2}$ s$^{-1}$.

Alternatively we can insert the explicit definition of
the Alfv\'en velocity (Eq.~5), $v_A = B_{np}/\sqrt{4 \pi m_p n_e} \approx
2 \times 10^{11} B_{np} n_e^{-1/2}$, which yields a modified
scaling law of 
\begin{equation}
	E_{MR2} 
        \approx {1 \over 4 \pi \sqrt{4 \pi m_p}}\ B_f^2\ B_{np}\ 
	L^2\ n_e^{-1/2}\ M_A\ \tau_f \ . 
\end{equation}
and a modified energy dissipation rate of 
\begin{equation}
	\varepsilon_{MR2} = {E_{MR2} \over V \tau_f}  
	= {E_{MR2} \over L^3\ \tau_f}  
        \approx {1 \over 4 \pi \sqrt{4 \pi m_p}}\ 
	B_f^2\ B_{np} L^{-1}\ n_e^{-1/2}\ M_A\ \tau_f \ . 
\end{equation}

For both models ($E_{MR1}$ and $E_{MR2}$), the obtained dissipated energies  
are much smaller than typical flare energies, found
to be in the range of $E_f\approx 10^{28}-10^{32}$ (Crosby et al.~1993).
Consequently, the Sweet-Parker reconnection rate is much too slow to 
explain the magnetic dissipation in solar flare events.
However, such small dissipated energies have been employed in the
interpretation of (undetected) nanoflares (Parker 1983) and of
(detected) smallest extreme-ultraviolet (EUV) flare-like
brightenings (Benz and Krucker 1999; Parnell and Jupp 2000;
Aschwanden et al.~2000), in the energy range of 
$E_f\approx 10^{24}-10^{26}$ erg.

\subsection{	Petschek Model		}

A much faster reconnection model was proposed by Petschek (1964), 
which involved reducing the size of the diffusion region to a very 
compact area $(\Delta \approx \delta)$ that is much shorter than 
the Sweet$-$Parker current sheet ($\Delta \gg \delta$) (Fig.~1, 
right panel). 
While the original (2-D) version of the Sweet-Parker model 
envisions a single large vertical current sheet located above a 
flare arcade (Fig. 2, left panel), the original Petschek model 
encompasses a much smaller current sheet with the 2-D geometry 
of an X-point. However, more realistic (3-D) models of the 
Petschek-type that account for sufficient reconnection flux invoke 
many small-scale current sheets that are formed by the tearing mode 
in coalescing magnetic islands with magnetic O-points and X- points 
(Fig. 2, right panel; Aschwanden 2002 and references therein).

Summaries of the Petschek model can be found 
in Priest (1982, p.~351), Jardine (1991), Priest \& Forbes 
(2000, p.~130), Treumann \& Baumjohann (1997, p.~148), and 
Tajima \& Shibata (2002, p.~225). Because the length of the current 
sheet is much shorter, the propagation time through the diffusion 
region is shorter and the reconnection process becomes faster.
However, in a given external area with size $L_e$ comparable with 
the length ${\Delta}$ of the Sweet$-$Parker current sheet, 
a much smaller fraction of the plasma flows through
the Petschek diffusion region with size ${\Delta}_P$, where finite 
resistivity $\sigma$ exists and field lines reconnect. Most of the 
inflowing plasma turns around outside the small diffusion region
and {\sl slow-mode shocks} arise where the abrupt flow speed
changes from $v_1$ to $v_2=v_A$ in the outflow region 
(Fig.~1, bottom panel). The shock waves represent an obstacle in 
the flow and thus are the main sites where inflowing magnetic 
energy is converted into heat and kinetic energy. Simple energy 
considerations show that inflowing kinetic energy is split up roughly in
equal parts into kinetic and thermal energy in the outflowing plasma 
(Priest \& Forbes 2000). Petschek (1964) estimated the maximum 
flow speed $v_e$ by assuming a magnetic potential field
in the inflow region and found that at large distance $L_e$ the 
external field $B_0(L_e)$ scales logarithmically with distance $L_e$,
\begin{equation}
        B_0(L_e) = B_0 \left[ 1 - {4 M_A \over \pi} \ln{ \left( { L_e 
	\over \Delta } \right) } \right]\ .
\end{equation}
Petschek (1964) estimated the maximum reconnection rate $M_A$ at a 
distance $L_e$ where the internal magnetic field is half of the 
external value (i.e., $B_0(L_e)=B_0/2$), which yields using Eq.~(14),
\begin{equation}
        M_A = {\pi \over 8 \ \ln{(  L_e / \Delta )} }  \approx
              {\pi \over 8 \ \ln{(R_{me})} } \ .
\end{equation}
So, the reconnection rate $M_A=v_1/v_A$ depends only 
logarithmically on the magnetic Reynolds number $R_{me} = 
L_e v_{Ae}/\eta$. Therefore, for coronal conditions,
where the magnetic Reynolds number is very high (i.e., 
$R_{me} \approx 10^8-10^{12}$), the Petschek reconnection rate is 
$M_A \approx 0.01-0.02$ according to Eq.~(15), yielding
an inflow speed of $v_1 \approx v_A\ M_A \approx 
10-20$ km s$^{-1}$ for typical coronal Alfv\'en speeds of 
$v_A\approx 1000$ km s$^{-1}$. Thus, the Petschek
reconnection rate is about three orders of magnitude 
faster than the Sweet$-$Parker reconnection rate.
The Petschek model can be expressed with the same formal 
scaling laws as the Sweet-Parker current sheet (Eq.~10 and 11), 
except for the expression of the Alfv\'enic Mach number (Eq. 15),
 which is substantially higher than in the Sweet-Parker current
 sheet model (Eq.~8).

MHD (hybrid) simulations have shown that magnetic reconnection
remains Alfv\'enic in a collisionless system even for very large
macroscopic scale lengths, yielding a universal constant for the
reconnection rate with an inflow velocity towards the X-line
around $v_1 \approx 0.1\ v_A$, which implies a Mach number of
$M_A=0.1$ (Shay et al.~1999).

The canonical volume definition $V(L)=L^3$ is subject to the
hydrostatic density structure of the solar corona, which 
can introduce a hydrostatic weighting bias in coronal temperature
measurements (Aschwanden and Nitta 2000). Since the vertical
electron mass density approximately follows an exponential
function with a scale height $\lambda(T_e)$ that depends linearly
on the temperature $T_e$, the volume of the diffusion region
$V_\lambda$ in a vertical current sheet can be modeled with a 
square area $A$ and a vertical scale height $\lambda$,
\begin{equation}
	V_\lambda(T) = A \lambda = L^2 \ \lambda \ , \qquad
	\lambda \approx 5 \times 10^9 \left( {T_e \over {\rm 1\ MK}} \right)\ 
	[{\rm cm}] \ ,
\end{equation}
which also affects the scaling laws. Therefore, the
model MR1 (Eq.~10) is modified to
\begin{equation}
	E_{MR3} = \int \left( {dE \over dt} \right) \ dt
        \approx {B_f^2 \over 4\pi}  L\ \lambda\ v_1     \ \tau_f
        \approx {1 \over 4\pi}\ B_f^2 L\ \lambda\ v_A\ M_A\ \tau_f \ , 
\end{equation}
and similarly model MR2 (Eq.~12) is modified to 
\begin{equation}
	E_{MR4} 
        \approx {1 \over 4 \pi \sqrt{4 \pi m_p}} \ 
	B_f^2\ B_{np}\ L\ n_e^{-1/2}\ \lambda\ M_A\ \tau_f \ . 
\end{equation}
Thus we specified four different versions of Petschek models, i.e.,
MR1 (Eq.~10), MR2 (Eq.~12), MR3 (Eq.~17), MR4 (Eq.~18), which
differ from each other in the assumptions made in the 
Alfv\'en velocity measurements ($v_A=const$ versus
$v_A \propto B_f/\sqrt{n_e}$) or different volume scaling laws
($V \propto L^3$ versus $V \propto L^2 \lambda$). 

\section{	OBSERVATIONAL DATA ANALYSIS 			}

\subsection{	Observations					}

We use the same data set of 173 solar flares presented in 
Paper I (Aschwanden et al.~2014) and 
Paper II (Aschwanden et al.~2015),  
which includes all M- and X-class flares observed with the SDO
(Pesnell et al.~2011) during the first 3.5 years of the mission
(2010 June 1 to 2014 January 31). Eliminating of data gaps or 
outliers leaves us with 167 events (or 162 correlations) 
for further analysis. This selection of events has a
heliographic longitude range of $[-45^\circ, +45^\circ]$, for which
magnetic field modeling can be faciliated without too severe
foreshortening effects near the solar limb. We use the 45-s
line-of-sight magnetograms from HMI/SDO and make use of all
coronal EUV channels of AIA (Lemen et al.~2012) onboard 
the {\sl Solar Dynamics Observatory (SDO)} (in the six wavelengths
94, 131, 171, 193, 211, 335 \ang ), which are sensitive to
strong iron lines in the temperature range of $T \approx 0.6-16$ MK. 
The spatial resolution is $\approx1.6"$ (0.6" pixels) for AIA, and
the pixel size of HMI is 0.5". The coronal magnetic field is
modeled by using the line-of-sight
magnetogram $B_z(x,y)$ from the {\sl Helioseismic and Magnetic
Imager (HMI)} (Scherrer et al.~2012) and (automatically detected)
projected loop coordinates $[x(s), y(s)]$ in each EUV wavelength of AIA.
A full 3-D magnetic field model ${\bf B}(x,y,z)$ is computed
for each time interval and flare with a cadence of 6 min,
where the total duration of a flare is defined
by the GOES flare start and end times. The size of
the computation box amounts to an area with a width and length
of 0.5 solar radii in the plane-of-sky, and an altitude range
of 0.2 solar radius. The total number of analyzed data
includes 2706 HMI images and 16,236 AIA images.

For the data analysis of this study, which is focused on the
magnetic reconnection rate, we require the following observables:
the mean nonpotential magnetic field strength $B_{np}$,
the mean magnetic field component $B_f=B_{\varphi}$ 
associated with the free energy or azimuthal field component,
the spatial length scale $L$ of the magnetic area in the flare region,
the mean electron density $n_e$, the flare duration $\tau_f$,
the free energy $E_f$, and the total dissipated magnetic
energy $E_{diss}$ during the flare duration. 

The magnetic parameters $B_{np}$,
$B_\varphi$, $E_f, E_{diss}$ are all computed
with the {\sl Vertical-Current Approximation Nonlinear
Force-Free Field (VCA-NLFFF)} code, as described in Paper I
and Paper IX. This magnetic field extrapolation code essentially
assumes vertical currents at flare locations that are associated
with sub-photospheric magnetic field concentrations (e.g., sunspots
and smaller magnetic features).  A major advantage of this code over 
traditional NLFFF codes is the capability to measure the 
current-driven twisting of coronal magnetic field lines, based 
on automated tracing of coronal loops in AIA images, which this 
way bypasses the non-force-freeness of the photospheric field.
 
The spatial scale $L$ is measured from the area $A=L^2$ of the 
(time-accumulated) azimuthal magnetic field, i.e., $B_\varphi(x,y)
 \ge 100$ G, after correction of projection effects (Paper I). 
The electron density $n_e$ is obtained from
a {\sl differential emission measure (DEM)} analysis, using 
the relationship of the total emission measure $EM$ with the
electron density, i.e., $EM = n_e^2 V \approx n_e^2 L^3$
(Paper II). The quantities $B_{\varphi}, L, n_e$ are determined
at the peak times of the flare emission measure, while the
parameters $E_f$ and $E_{diss}$ are integrated over
the flare duration $\tau$. The minimum, maximum, mean and
standard deviation, and median values are listed in Table 1
and histogrammed in Fig.~3.

\subsection{	Magnetic Reconnection Scaling Law Tests	}

The total dissipated energy $E_{diss}$ in a flare is a
fundamental parameter, for which a scaling law has been
considered in terms of the Sweet-Parker model (Section 2.1) 
and the Petschek model (Section 2.2), according to Eq.~(10),
where an Alfv\'en Mach number of $M_A \approx 10^{-5}$ 
is typical for the Sweet-Parker model (Eq.~8), and
a value of $M_A \approx 10^{-1}$ for the Petschek model
(Eq.~15). We plot the theoretically predicted magnetic 
dissipation energies $E_{MR1}^{pred}$ of the first model 
(Eq.~10) as a function of the observed dissipation
energies $E_{MR1}^{diss}$ in Fig.~4a. The theoretical 
model is obviously related to the observed values.
We measure three different criteria to quantify this
relationship: (i) the cross-corrletagion coefficient
\begin{equation}
        CCC = { \sum_i E_{pred} E_{diss} \over
                \sum_i E_{pred} \times \sum_i E_{diss} } \ ,
\end{equation}
between the observed $E_{diss}$ and the scaling-law
predicted theoretical values $E_{pred}$; (ii)
the slope $p$ of the linear regression fit; and (iii)
the median energy ratio $q_E=E_{pred}/E_{obs}$.
These three criteria have the values of
$CCC=0.79$, $p=1.35$, and $q_E=0.68$ for model MR1 (Fig.~4a), 
so it indicates a positive correlation, but does not
match exactly the linear regression slope $p$ and
median energy ratio $q_E$.

In model MR2 (Fig.~4b) we use the actually observed 
values $B_{np}$ and $n_e$ in the calculation of the Alfv\'en
speed, while we used the constant value $v_A=1700$ 
km s$^{-1}$ in model [MR1], which is the median value of all events.
The results are not much different in the second model
(MR2), except for a worse linear regression slope of
$p=1.52$, likely to be caused by uncertainties of the
parameters $B_{np}$ and $n_e$.

In the models MR3 (Fig.~4c) and MR4 (Fig.~4d) we use 
the hydrodstatic volume scaling $V=L^2 \lambda$, 
in contrast to the standard volume cubes $V=L^3$ 
used in the first two models (MR1, MR2). 
Interestingly, the linear regression slope $p$
improves to a value of $p=1.07$, which is close to
the theoretically expected value $p=1.0$ in the case
of a matching scaling law. Apparently, consideration
of the hydrostatic scale height is a necessary
ingredient to improve the empirical scaling law
for magnetic reconnection, which has been ignored
in previous studies.

What do these results tell us? If we accept
the model MR3 as the best match 
to a magnetic reconnection model, we learn (i)
that the high cross-correlation coefficient of
$CCC=0.73$ implies that the underlying scaling
law $E_{diss} \propto B_f^2\ L\ v_A$ (Eq.~17) 
renders the most likely functional relationship,
(ii) that an Alfv\'en Mach number of $M_A \approx 0.1$ 
predicts the correct magnitude of an energy
dissipation event, and (iii) that the
hydrostatic scaling of the diffusion region
has to be taken into account to obtain
accurate mean electron densities.

As a caveat, we have to consider the ambiguity
of the values in the product $\lambda M_A$
(Eq.~17) in the scaling law of model MR3.
The exponential electron density scale height
scales linearly with the temperature,
$\lambda \propto T_e$, which is not explicitly
measured here. The Alfv\'en Mach number $M_A$
is another parameter that is not known in 
our data set. Thus, the product of these
two quantities is ambiguous. In the models
MR1-MR4 discussed here we assumed a coronal
temperature of $T_e=1.0$ MK and a Mach number
of $M_A=0.1$. Theoretically, a reciprocal
choice of $T_e=2.0$ MK and $M_A=0.05$ would
yield an identical product. However, it has been 
argued that the inflows into the diffusion
region of a magnetic reconnection process
are expected to be of coronal temperatures
($T_e \approx 1-2$ MK), rather than of flare plasma
temperatures ($T_e \approx 5-25$ MK), because the
flowing plasma would be supplied from the corona 
surrounding the flare (Hara and Ichimoto 1996).
 	
\subsection{	Alfv\'en Mach Number		}

Adopting the best-fitting model MR3 (Eq.~17) we obtain
a simple expression for inferring the Alfv\'en Mach Number
$M_A$ from the observables ($B_f$, $L$, $\lambda$, $v_A$, $\tau_f$),
\begin{equation}
        M_A={ 4 \pi E_{diss} \over B_f^2\ L\ \lambda\ v_A\ \tau_f} \ ,
\end{equation}
where the mean free energy field strength $B_f=B_\varphi$ 
and dissipated energy $E_{diss}$ are
obtained from the nonlinear force-free field solution 
${\bf B}_f$ averaged over the volume of the computation box,
$L=\sqrt{A}$ is the length scale obtained from the magnetic
flare area $A$ above some threshold (e.g., $B > 100$ G),
$v_A = B_{np}/\sqrt{4 \pi m_p n_e}$ is the Alfv\'en velocity
measured in the reconnection outflow, given by the mean
nonpotential force-free solution ${\bf B}_{np}$, 
the electron density $n_e$ in the reconnection outflow, and the
flare duration $\tau_f$, for instance from the GOES flare
start and end time. We show the results of the so obtained
Alfv\'en Mach numbers $M_A$ in Fig.~5, juxtaposed to other
previous measurements. Note that most observational results of 
Mach numbers were inferred from velocity measurements of 
reconnection inflows or outflows (e.g., with Yohkoh/SXT),
rather than from nonlinear force-free field solutions
as done here. In this sense we offer a new method to
determine the Alfv\'en Mach number, based on observations
of the magnetic field (HMI/SDO) and EUV images (AIA/SDO).

The Alfv\'enic Mach number is defined as $M_A=v1/v_A$, 
where the Alfven speed is defined by $v_A=B_f/\sqrt{n_e}$, 
with $B_f$ the mean magnetic energy in the active region 
and $n_e$ the mean electron density. Since the soft X-ray 
emission measure, $EM=n_e^2\ V$ is a good proxy for the soft 
X-ray flux $F_{SXR}$, we expect that the Alfv\'enic Mach 
number $M_A$ scales slightly positive with the emission 
measure or soft X-ray flux, i.e., $M_A \propto EM^{1/4} 
\propto F_{SXR}^{1/4}$.

\section{		DISCUSSION 			}

\subsection{	Evidence for Magnetic Reconnection Inflows 	}

All magnetic reconnection models require inflows into the sides 
of the current sheet and predict near-Alfv\'enic outflows. 
In solar flares, however, it turned out to be extremely 
difficult to detect reconnection inflows, either because of 
low density, low contrast, slow inflow speed, incomplete 
temperature coverage, or projection effects (Hara and Ichimoto 1996). 
In the famous ``candle-light'' flare event observed with Yohkoh/SXT
(Soft X-ray Telescope), inflow speeds of $v_{in} \approx 56$
km s$^{-1}$ ($M_A \approx 0.07$) and outflow speeds of 
$v_{out} \approx 800$ km s$^{-1}$ ($M_A \approx 1$) were estimated 
based on hydrodynamic modeling arguments (Tsuneta (1996).
During the Yohkoh era, only one study has been published that shows 
direct evidence for reconnection inflows in a solar flare,
where an inflow speed of $v_{in} \lapprox 5$ km s$^{-1}$ was
measured with EIT/SOHO and a reconnection rate of 
$M_A \approx 0.001-0.03$ was estimated
(Yokoyama et al.~2001), consistent with the range of other estimates,
$M_A \approx 0.001-0.01$ (Isobe et al.~2002), see also Fig.~5 for
comparisons. 

Direct measurements of reconnection outflows are not readily observed either. 
The first evidence of high-speed downflows above flare loops is believed to 
be observed by Yohkoh/ SXT during the 1999-Jan-20 flare, showing dark 
voids flowing down from the cusps, with speeds of $v_{out}\approx 100-200$ 
km s$^{-1}$ (McKenzie \& Hudson 1999; McKenzie 2000),
about an order of magnitude slower than expected for coronal Alfv\'en speeds.
The downward outflows hit a high-temperature ($T\approx 15-20$ MK) region,
which might be evidence of the fast-mode shock (Tsuneta 1997a, 1997b),
sandwiched between the two ridges of the slow-mode shock (Tsuneta 1996).
Blueshifted Fe XXV lines were also interpreted as direct outflows from 
Petschek-type reconnection regions (Lin et al.~1996).

\subsection{		Alfv\'en Mach Number 			}

A key parameter to characterize the dynamics and geometry of the 
diffusion region of a magnetic reconnection process is the
Alfv\'en Mach number $M_A$, which is the ratio of the sideward
inflow velocity $v_1$ to the vertical outflow speed $v_2$ (Eq.~8).
In the Sweet-Parker model (Sweet 1958; Parker 1963), 
a very low value of $M_A=1/\sqrt{S} \approx 10^{-5}$ was adopted, 
due to the high value of the Lundquist number $S \approx 10^{10}$ 
adopted for the solar corona. Petschek (1964) derived a much
higher value of $M_A \approx 0.01-0.02$, due to the logarithmic
dependence on the Reynolds number. Further theoretical studies
that include the external region for Petschek's mechanism in
greater detail revealed upper limits at approximately 20\%
of Petschek's value (Roberts and Priest 1975).
This trend continued with numerical MHD simulations, 
leading to high values of up to $M_A \approx 0.1$ (Shay et al.~1999).
For a juxtaposition of different Alfv\'en Mach numbers see Fig.~5.
The Petschek's reconnection mechanism is considered to be a 
particular solution of the MHD equations that apply only 
when special conditions are met (Forbes 2001).
Alternatively to MHD simulations, two-fluid concepts were 
employed, where ions and electrons have different temperatures,
which splits the reconnection outflow into multiple streaming
channels (Longcope and Bradshaw 2010).

\subsection{		Flare Ribbon Motion 		}

Magnetic reconnection is often modeled as a quasi-stationary
steady-state scenario, operating in an equilibrium between
lateral inflows into a vertical current sheet, and 
reconnection outflows in upward and downward direction.
There is also evidence for dynamic time evolution,
where the footpoint ribbons systematically move apart, 
while the centroid of the magnetic reconnection X-point
rises upward, as a consequence of the self-similar 
reconnection geometry. Such separation motion of 
soft X-ray and EUV flare ribbons have been observed 
in many flares (e.g., Tsuneta et al.~1992; Fletcher and Hudson 2001;
Krucker et al.~2003; Asai et al.~2004; Sui et al.~2004;
Veronig et al.~2006; Vrsnak et al.~2006; Temmer et al.~2007).
Another dynamics of flare ribbons is the elongation motion
along the polarity inversion line (Qiu et al.~2017).

The related energy
dissipation rate can be written as a product of the Poynting
flux into the reconnection region, $S=B_c v_i/(4 \pi)$,
and the area of the reconnection region, A,
\begin{equation}
	{dE \over dt} = S A = 2 {B_c^2 \over 4\pi} v_i A 
	              = 2 {B_c^2 \over 4\pi} v_A\ M_A\ A \ ,
\end{equation} 
where $B_c$ is the mean magnetic field strength
in the corona, and $v_i$ is the inflow velocity into the
reconnection region (Asai et al.~2004). Strictly speaking,
the coronal magnetic field strength should be estimated
with the magnetic field component that is associated with
the free energy, $B_c \approx B_f$ (see Eq.~11).
Nevertheless, this model predicts a decreasing energy
dissipation rate $E(t)$ when reconnection proceeds to
higher altitudes, where $B_c(t)$ and $v_i(t)$ decrease
with progressive height. More generally, measuring the
time-dependent parameters [$B_f(t), v_A(t), M_A(t)$]
from observations and integrating the time-dependent
parameters over the flare duration would lead to more 
accurate estimates of the flare-dissipated energies,
rather than multiplying with the flare duration $\tau_f$
(as done in Eq.~10).

\subsection{		Scaling Law Parameterization 		}

The long-standing coronal heating problem, which is related
to the coronal flare heating problem, has been characterized
with scaling laws that involve combinations of physical 
parameters raised to some power. A variety of 22 different
scaling laws of the form $\varepsilon_{diss} \propto B^a L^b \rho^c
v^d R_c$ has been presented in Mandrini et al.~(2000), where
$B$ is the average coronal field strength (computed with
potential, linear force-free, and magnetostatic extrapolation
models), $L$ a characteristic length scale of coronal loops,
$\rho = m_p n_e$ the mean mass density, $v$ is the transverse
velocity at the base of the corona, and $R_c$ is a characteristic
horizontal length scale for the magnetic field or flow field. 
Based on a previous study, a universal scaling law of
$<B> \propto L^\delta$ with a mean slope of $\delta=
-0.88 \pm 0.3$ has been claimed (Klimchuk and Porter 1995), 
but the power law slope $\delta$ was found to be different in the
near-field and far-field of a magnetic dipole field, fitting
an empirical function of $<B>(L) = c_1 + c_2\ log(L) + 
C_3/2\ log(L^2+S^2)$ (Mandrini et al.~2000). Although we deal
with the same problem here, our present study 
differs in a number of assumptions: (i) The dissipated magnetic
energy cannot exceed the free energy, $E_{diss} \le E_f$, and
thus the associated magnetic field component is 
${\bf B}_f = {\bf B}_{np} - {\bf B}_p$, rather than the
average coronal field strength $<B> = <{\bf B}_{np}>$ used in
Mandrini et al.~(2000); (ii) The universal scaling law
$<B> \propto L^{\delta}$ claimed by Klimchuk and Porter (1995)
does not hold over the near-field range, 
and the fitted empirical function (with logarithmic terms)
is unphysical; (iii) The electron density scale height $\lambda$
of the hydrostatic corona is neglected in Mandrini et al.~(2000); 
(iv) Important physical parameters are missing in  
scaling laws given by Mandrini et al.~(2000), 
which are relevant in most of the included magnetic reconnection 
scenarios, such as the inflow speed $v_1$, the Alfv\'en speed $v_A$, 
or the Alfv\'en Mach number $M_A$.

A related method to test scaling laws of coronal heating
mechanisms was attempted with full-Sun visualizations by
Schrijver et al.~(2004). Images of the full Sun are simulated
in various wavelengths by populating the corona with a total
of 50,000 coronal loops, each one characterized with a 
quasi-static loop top temperature $T_e$, footpoint electron 
density $n_e$, and steady-state heating rate $\varepsilon_{heat}$.
The best match between the observed and simulated soft X-ray
and EUV emission of active regions is found for the scaling
law of the Poynting flux $P_{heat} \propto B^{1.0\pm0.3}
L^{-1.0 \pm 0.5}$ (in units of [erg cm$^{-2}$ s$^{-1}$]), 
which corresponds to a volumetric heating rate of
$\varepsilon_{diss} \propto P_{heat} / L \propto B^{1.0\pm0.3} 
L^{-2\pm 1}$. 
This result is not consistent with any of our 4 models
(MR1: Eq.~10, MR2: Eq.~12, MR3: Eq.~17, MR4: (Eq.~18). 
We suspect that similar factors contribute to this
discrepancy as discussed above in the study of Mandrini 
et al.~(2000), such as the invalidity of the scaling of
$B \propto L^\delta$, the proper distinction between the
potential, the nonpotential, and free energy in the
scaling laws containing the magnetic field $B$, as well as 
the neglect of physical parameters relevant in magnetic 
reconnection scenarios (inflow speed $v_1$, Alfv\'en speed 
$v_A$, and Alfv\'en Mach number $M_A$).

While we focused on magnetic scaling laws of flare energies 
in this study, there exist additional complementary flare scaling 
laws that involve thermal energies and emission measures (Papers II, V), 
such as the Rosner-Tucker-Vaiana (Rosner et al.~1978) and the 
Shibata-Yokoyama scaling law (Shibata and Yokoyama 1999), subject of future 
studies on heating rates and thermal energies.” 

\section{		CONCLUSIONS 			}

In this study we focus on testing scaling laws of magnetic
flare energies, which is related to magnetic scaling laws
responsible for coronal heating, as investigated in previous 
studies (Mandarini et al.~2000; Schrijver et al.~2004).  
The most basic magnetic scaling law is the Sweet-Parker
model for magnetic reconnection, which has been improved
into a more realistic model by Petschek (1964). Although
the physical parameters of these theoretical models are 
well-defined, the application of these scaling laws to
observables is not trivial, because of spatio-temporal
averaging of flare regions, their restricted observability 
in 2-D, and the reconstructability of their 3-D geometry.
We pursue a new approach with a nonlinear force-free
field code that provides the 3-D magnetic structure
and quantitative measurements of magnetic parameters,
such as the dissipated magnetic energy, the potential,
non-potential and free energies, 
the Alfv\'en velocity, the reconnection rate,
and the Alfv\'en Mach number. We are using identical data
sets as used previously in Paper I and II on the global
energetics of flares and CMEs (Aschwanden et al.~2014, 2015),
which includes only M-class and X-class flares.
In the following we summarize the conclusions of this study:

\begin{enumerate}

\item{The Sweet-Parker scaling law provides a suitable formalism
for the energetics of a basic magnetic reconnection process, 
but uses an unrealistic Alfv\'en Mach number based on the high
magnetic Reynolds number of order $S \approx 10^{10}$, which
predicts extremely thin current sheets with widths $\delta
\approx 10^{-5}\ L$ (that are not stable and would break up 
due to the tearing mode), and have a too slow reconnection rate to
explain solar flares.}

\item{The Petschek model assumes a potential field in the inflows
of the diffusive magnetic reconnection region and finds a
logarithmic dependence on the magnetic Reynolds number, which
predicts a large reduction in the size of the diffusion region,
a much faster reconnection rate (about 3 orders of magnitude
faster than the Sweet-Parker model), and an Alfv\'en Mach number of
$M_A \approx 0.1$ (Fig.~5).}

\item{For the magnetically dissipated energy $E_{diss}$ in flares we
find a good correlation for the scaling law 
$E_{diss} = (1/4\pi) B_f^2 L^2 v_A M_A \tau_f$, with a cross-correlation
coefficient of $CCC=0.79$ and a mean energy ratio of $q_E=0.73$.
The median Alfv\'en velocity is $v_A=1700$ km s$^{-1}$ in this model MR1.
Inserting the flare-averaged (nonpotential) magnetic field strengths
$B_{np}$ and electron densities $n_e$ into the expression of the
Alfv\'en velocity $v_A$, we find a similar good correlation (in model MR2).}

\item{Previous estimates of the reconnection inflow rate used a volume
scaling of $V \propto L^3$, which implies a reconnection outflow rate of
$dV/dt = L^2 v_A$. Since the electron density distribution of the solar
corona is approximately hydrostatic, i.e., $n_e(h) \propto e^{-h/\lambda}$,
a volume scaling of $V \propto L^2 \lambda$ is more realistic, which
implies a reconnection outflow rate of $dV/dt = L\ \lambda\ v_A$. With
this hydrostatic correction we obtain a best-fitting model (MR3) that
has a cross-correlation coefficient of $CCC=0.73$, a mean energy ratio
of $q_E=0.96$, as well as a slope $p=1.07$ of the linear regression fit,
which is close to the ideal value of $p=1.0$ for a perfect match.
Since the inflow originates from outside the flare core, the density
scale height $\lambda$ corresponds to the average coronal temperature
of $T_e \approx 1.0$ MK. This optimization changes the scaling law to
$E_{diss} = (1/4\pi) B_f^2\ L\ \lambda\ v_A\ M_A\ \tau_f$.}

\item{Measurements of magnetic reconnection inflows have been sparse,
due to the low densities, low contrast, low inflow speed, incomplete
temperature coverage, or projection effects. Such measurements have been
carried out in Yohkoh/SXT data, where the lateral inflow speed was
inferred from the spatial displacement of plasma in the cusp regions.
Here we present a new method that can corroborate the directly observable 
projected velocity, or future spectroscopic velocity measurements.}

\item{There are numerous observations and  measurements of flare ribbons,
which contain information on the time-dependent energy dissipation 
$E_{diss}(t)$ in magnetic reconnection sites, which allows us to
study the time evolution of magnetic reconnection processes in flares.}

\item{Based on the best-fitting magnetic energy scaling law 
evaluated here, which requires a discrimination between the magnetic
free energy $E_f$ and the nonpotential energy $E_{np}$, and of the
related magnetic field strengths $B_f$ and $B_{np}$, we conclude
that hypothetical scaling laws with a single component $B$ represent
oversimplified approaches that cannot retrieve physically correct
scaling laws. Moreover, the scaling $B(L) \propto L^\delta$ is
invalid, because the power law coefficient differs in the near-field
and far-field range of a magnetic dipole.}

\end{enumerate}

In summary, this study quantifies the scaling law of magnetic
reconnection with a number of magnetic parameters ($B_f, B_{np},
v_A, M_A$) that can be used to predict the total magnetic energy 
$E_{diss}$ dissipated in solar flares, which provides not only a 
deeper physical understanding of the underlying magnetic 
reconnection process, but serves also for statistical predictions 
of flare energies that are needed in space weather forecasting.

\acknowledgements
Part of the work was supported by NASA contract NNG 04EA00C of the 
SDO/AIA instrument and the NASA STEREO mission under NRL contract 
N00173-02-C-2035.

\clearpage

\section*{REFERENCES}

\def\ref#1{\par\noindent\hangindent1cm {#1}}

\ref{Asai, A., Yokoyama, T., Shimojo, M., Masuda, S., Kurokawa, H., 
	and Shibata, K. 2004, ApJ 611, 557}
\ref{Aschwanden, M.J., Benz, A.O., Dennis, B.R. and Gaizauskas, V.
 	1993, ApJ 416, 857}
\ref{Aschwanden, M.J., Benz, A.O., Dennis, B.R., and Schwartz, R.A.
 	1995, ApJ 455, 347}
\ref{Aschwanden, M.J., Wills, M.J., Hudson, H.S., Kosugi, T., 
	and Schwartz, R.A. 1996, ApJ 468, 398}
\ref{Aschwanden, M.J., Tarbell, T., Nightingale, R., Schrijver, C.J., 
	Title, A., Kankelborg, C.C., Martens, P.C.H., and Warren, H.P.
 	2000, ApJ 535, 1047}
\ref{Aschwanden, M.J. and Nitta, N.  2000, ApJ 535, L59}
\ref{Aschwanden, M.J.  2002, SSRv 101, 1}
\ref{Aschwanden, M.J.  2004, Praxis and Springer, New York, 
	ISBN 3-540-22321-5, First Edition, hardbound issue, 842p
 	Physics of the Solar Corona - An Introduction (1st Edition)}
\ref{Aschwanden, M.J., Xu, Y., and Jing, J. 2014, ApJ 797, 50, (Paper I)}
\ref{Aschwanden, M.J., Boerner,P., Ryan, D., Caspi, A., McTiernan, J.M., 
	and Warren, H.P. 2015, ApJ 802:53, (Paper II)} 
\ref{Aschwanden, M.J. 2019, ApJ 885:49, (Paper IX)} 
\ref{Benz, A.O. and Krucker S., 1999, AA 341, 286}
\ref{Biernat, H.K., Heyn, M.F., and Semenov, V.S.  1987, JGR 92/A4, 3392}
\ref{B\"uchner, J.  2006, SSRv 124, 345}
\ref{Carmichael, H. 1964, in proc. {\sl The Physics of Solar Flares}, 
	NASA Special Publication 50, (ed. W.N. Hess), NASA, Washington DC, p.451.}
\ref{Crosby, N.B., Aschwanden, M.J., and Dennis,B.R.
 	1993, SoPh 143, 275}
\ref{Fletcher, L. and Hudson, H.S. 2001, SoPh 204, 71}
\ref{Forbes, T.G. and Acton, L.W.
 	1996, ApJ 459, 330}
\ref{Forbes, T.G. 2001, Earth, Planets and Space 53, 423}
\ref{Gosling, J.T., Eriksson, S., Skoug, R.M., McComas, D.J. and
	Forsyth, R.J.  2006a, ApJ 644, 613}
\ref{Gosling, J.T., Eriksson, S., and Schwenn, R. 
	2006b, JRL, 111, A10, A10102}
\ref{Hara, H. and Ichimoto, K.
 	1996, in Proc. {\sl Magnetic Reconnection in the Solar Atmosphere}, 
	ASP Conf. 111, 183}
\ref{Hirayama, T. 1974, SoPh 34, 323}
\ref{Isobe, H., Yokoyama, T., Shimojo, M., Morimoto, T., Kozu, H., Eto, S., 
	Narukage, N., and Shibata, K.
 	2002, ApJ 566, 528}
\ref{Jardine, M. 1991, in {\sl Mechanisms of Chromospheric and Coronal Heating},
        (eds.~P. Ulmschneider, E.R. Priest, and R. Rosner), 
	Springer, Berlin, p.588.}
\ref{Klimchuk, J.A. and Porter, L.J. 1995, Nature, 377, 131}
\ref{Kopp, R.A. and Pneuman, G.W. 1976, SoPh 50, 85}
\ref{Krucker, S., Hurford, G.J., and Lin, R.P. 2003, ApJ 595, L103}
\ref{Kulsrud, R.M.  2001, Earth, Planets and Space 53, 417}
\ref{Lemen, J.R., Title, A.M., Akin, D.J., Boerner, P.F., Chou, C., Drake, J.F., 
	Duncan, D.W., Edwards, C.G., et al.  2012, SoPh 275, 17.}
\ref{Lin, J., Martin, R., and Wu, N. 1996, AA 311, 1015}
\ref{Mandrini, C.H., Demoulin, P., and Klimchuk, J.A.  2000, ApJ 530, 999}
\ref{Masuda, S., Kosugi, T., Hara, H., Tsuneta, S., and Ogawara, Y.
 	1994 Nature 371/6497, 495.}
\ref{McKenzie, D.E. and Hudson, H.S. 1999, ApJ 519, L93}
\ref{McKenzie, D.E. 2000, SoPh 195, 381}
\ref{Longcope, D.W. and Bradshaw, S.J. 2010, ApJ 718, 1491}
\ref{Parker, E.N.  1963, ApJS 8, 177}
\ref{Parker, E.N. 1983, ApJ 264, 642}
\ref{Parker, E.N. 1988, ApJ 330, 474}
\ref{Parnell, C.E. and Jupp, P.E. 2000, ApJ 529, 554}
\ref{Pesnell, W.D., Thompson, B.J., and Chamberlin, P.C. 2011, SoPh 275, 3}
\ref{Petschek, H.E. 1964, in "The Physics of Solar Flares", Proc. AAS-NASA 
	Symposium, (ed. W.N.Hess), Washington DC, p.425}
\ref{Priest, E.R. 1982 {\sl Solar Magnetohyrdodynamics} Geophysics and 
	Astrophysics Monographs 
	Volume 21, D. Reidel Publishing Company, Dordrecht}
\ref{Priest, E.R. and Forbes,T. 2000, 
 	Cambridge: Cambridge University Press,
 	Magnetic reconnection (MHD Theory and Applications)}
\ref{Qiu, J., Longcope, D.W., Cassak, P.A., and Priest, E.R.
 	2017, ApJ 838, 17}
\ref{Roberts, B. and Priest, E.R. 
	1975, J. Plasma Physics 14/3, 417}
\ref{Rosner, R., Tucker, W.H., and Vaiana, G.S. 1978, ApJ 220, 643} 
\ref{Sakao, T. 1994, PhD Thesis, University of Tokyo}
\ref{Sakao, T., Kosugi, T., and Masuda, S.
 	1998, in Proc. {\sl Observational Plasma Astrophysics: Five Years 
	of Yohkoh and Beyond}, (eds. Watanabe, T., Kosugi, T. and Sterling, A.C.), 
	Astrophysics and Space Science Library Vol. 229, p.273-284.
	Kluwer Academic Publishers, Dordrecht, Netherlands}
\ref{Scherrer, P.H., Schou, J., Bush, R.I., Kosovichev, A.G., Bogart, R.S., 
	Hoeksema, J. T., Liu, Y., Duvall, T.L., et al.
 	2012, SoPh 275, 207}
\ref{Schrijver, C.J., Sandman, A.W., Aschwanden, M.J., and DeRosa, M.L.
 	2004, ApJ 615, 512}
\ref{Shay, M.A., Drake, J.F., Rogers, B.N., and Denton, R.E.
 	1999, GRL 26, 2163}
\ref{Shibata, K., Masuda, S., Shimojo, M., Hara, H., Yokoyama, T., 
	Tsuneta, S., Kosugi, T., and Ogawara, Y. 1995, ApJ 451, L83}
\ref{Shibata, K. and Yokoyama, T. 1999, ApJ 526, L49} 
\ref{Sturrock, P.A. 1966, Nature 5050, 695}
\ref{Sui, L. and Holman, G.D. 2003, ApJ 596, L251}
\ref{Sui, L., Holman, G.D., and Dennis,B.R. 2004, ApJ 612, 546}
\ref{Svestka, Z., Fontenla, J.M., Machado, M.E., Martin, S.F., Neidig, D., 
	Poletto, G. 1987, SoPh 108, 237}
\ref{Sweet, P.A. 
 	1958, in "Electromagnetic Phenomena in Cosmic Physics", 
	International Astronomical Union Symposium 6, (ed. B.Lehnert), 
	Cambridge University Press, New York, p.123}
\ref{Tajima, T. and Shibata, K.
 	2002, {\sl Plasma Astrophysics},
	Perseus Publishing, Cambridge, Massachusetts}
\ref{Temmer, M., Veronig, A.M., Vrsnak, B., and Miklenic, C.  2007, ApJ 654:665}
\ref{Treumann, R.A. and Baumjohann,W.  1997, {\sl Advanced space plasma physics},
 	Imperial College London.}
\ref{Tsuneta, S., Hara, H., Shimizu, T., Acton, L.W., Strong, K.T., 
	Hudson, H.S., and Ogawara, Y. 1992, PASJ 44, L63}
\ref{Tsuneta, S.  1996, ApJ 456, 840}
\ref{Tsuneta, S., Masuda, S., Kosugi, T., and Sato, J. 1997a, ApJ 478, 787}
\ref{Tsuneta, S. 1997b, ApJ 483, 507}
\ref{Uzdensky, D.A. 2003, ApJ 587:450}
\ref{Veronig, A.M., Karlicky, M., Vrsnak, B., Temmer, M., Magdalenic, J., 
	Dennis, B.R., Otruba, W., and Poetzi, W. 2006, AA 446, 675}
\ref{Vrsnak, B., Temmer, M., Veronig, A., Karlicky, M., and Lin, J.
 	2006, SoPh 234, 273}
\ref{Yokoyama, T., Akita, K., Morimoto, T., Inoue, K., and Newmark,J.
 	2001, ApJ 546, L69}

\clearpage


\begin{table}[th]
\tabletypesize{\normalsize}
\caption{Observational parameters of $N=162$ M- and X-class flare events analyzed
in Aschwanden, Xu, and Jing (2014).}
\medskip
\begin{tabular}{lrrrr}
\hline
				& Minimum               & Maximum 	& Mean  	& Median \\
\hline
\hline
Free energy $E_f$ [10$^{30}$ erg] 
			        			& 0.4      & 950        & 120$\pm$160   & 64    \\
Dissipated magnetic energy $E_{diss}$ [10$^{30}$ erg]   & 1.5	   & 1500	& 180$\pm$250   & 110	\\
Nonpotential magnetic field strength $B_{np}$ [G]	& 80 	   & 900        & 270$\pm$160 	& 240 	\\
Free energy field strength $B_f=B_\varphi$ [G] 		& 14       & 390        & 80$\pm$60 	& 70 	\\
Length scale $L$ [Mm]					& 13 	   & 240        & 40$\pm$25	& 34 	\\
Electron density $n_e [10^{10}$ cm$^{-3}$]  		& 3	   & 56		& 10$\pm$7      & 9     \\
Alfv\'en velocity $v_A$ [km s$^{-1}$] 			& 380      & 8000	& 2100$\pm$1300 & 1700   \\
Flare duration $\tau_f$ [s]     			& 290	   & 15,000	& 1500$\pm$1800 & 900   \\
\hline
\end{tabular}
\end{table}

\bigskip


\begin{table}[h]
\tabletypesize{\normalsize}
\caption{Four versions of the Petschek model (MR1, MR2, MR3, MR4)
applied to 162 M- and X-class flare events analyzed in Aschwanden, 
Xu, and Jing (2014). The results are given by the power exponent $p$ 
of the scaling law, the energy ratio $q_E=E_{diss}/E_{obs}$ (computed 
for an Alfv\'en Mach number of $M_A=0.1$, and the cross-correlation 
coefficient $CCC$ between the theoretically scaling law and observed 
energy values.}
\medskip
\begin{tabular}{lllllll}
\hline
Model		& Observed      & Alfv\'en velocity,   & Volume  & Power    & Energy   & Cross-Correlation \\
		& energy	&                    & scaling & exponent & ratio	& coefficient	 \\
	        & $E$           & $v_A$	             & $V$     & $p$      & $q_E$    & CCC		 \\  
\hline
\hline
Petschek MR1	& $E_{diss}$	& $v_A=1700$ km  s${-1}$   & $V=L^3$         & $1.35$ & $0.68$ & 0.79      \\
Petschek MR2	& $E_{diss}$	& $v_A(B,n_e)$             & $V=L^3$         & $1.52$ & $0.64$ & 0.73      \\
Petschek MR3	& $E_{diss}$	& $v_A=1700$ km  s$^{-1}$  & $V=L^2 \lambda$ & $1.07$ & $0.96$ & 0.73      \\
Petschek MR4	& $E_{diss}$	& $v_A(B,n_e)$             & $V=L^2 \lambda$ & $1.23$ & $0.94$ & 0.67      \\
\hline
\end{tabular}
\end{table}

\clearpage


\begin{figure}
\centerline{\includegraphics[width=0.7\textwidth]{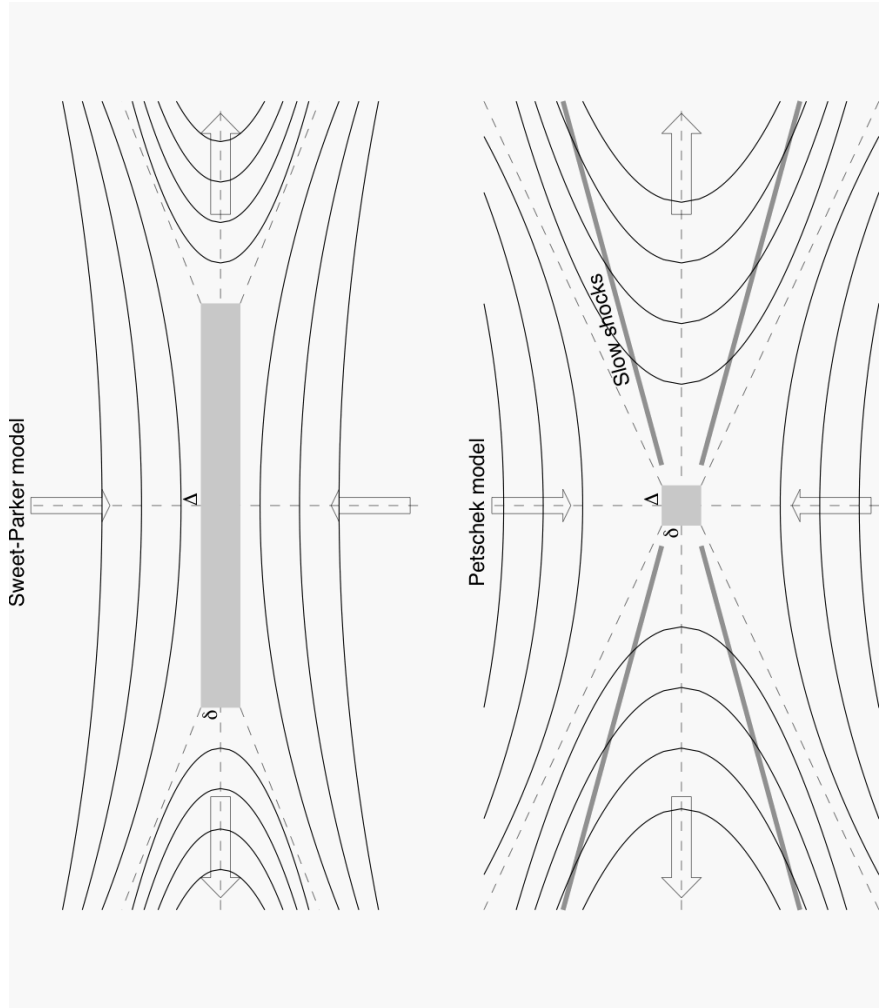}}
\caption{Geometry of the Sweet$-$Parker (top) and Petschek 
reconnection model (bottom).  The geometry of the diffusion 
region (grey box) is a long thin sheet ($\Delta \gg \delta$)
in the Sweet$-$Parker model, but much more compact ($\Delta 
\approx \delta$) in the Petschek model. The Petschek model 
also considers slow-mode MHD shocks in the outflow region.}
\end{figure}

\begin{figure}
\centerline{\includegraphics[width=1.0\textwidth]{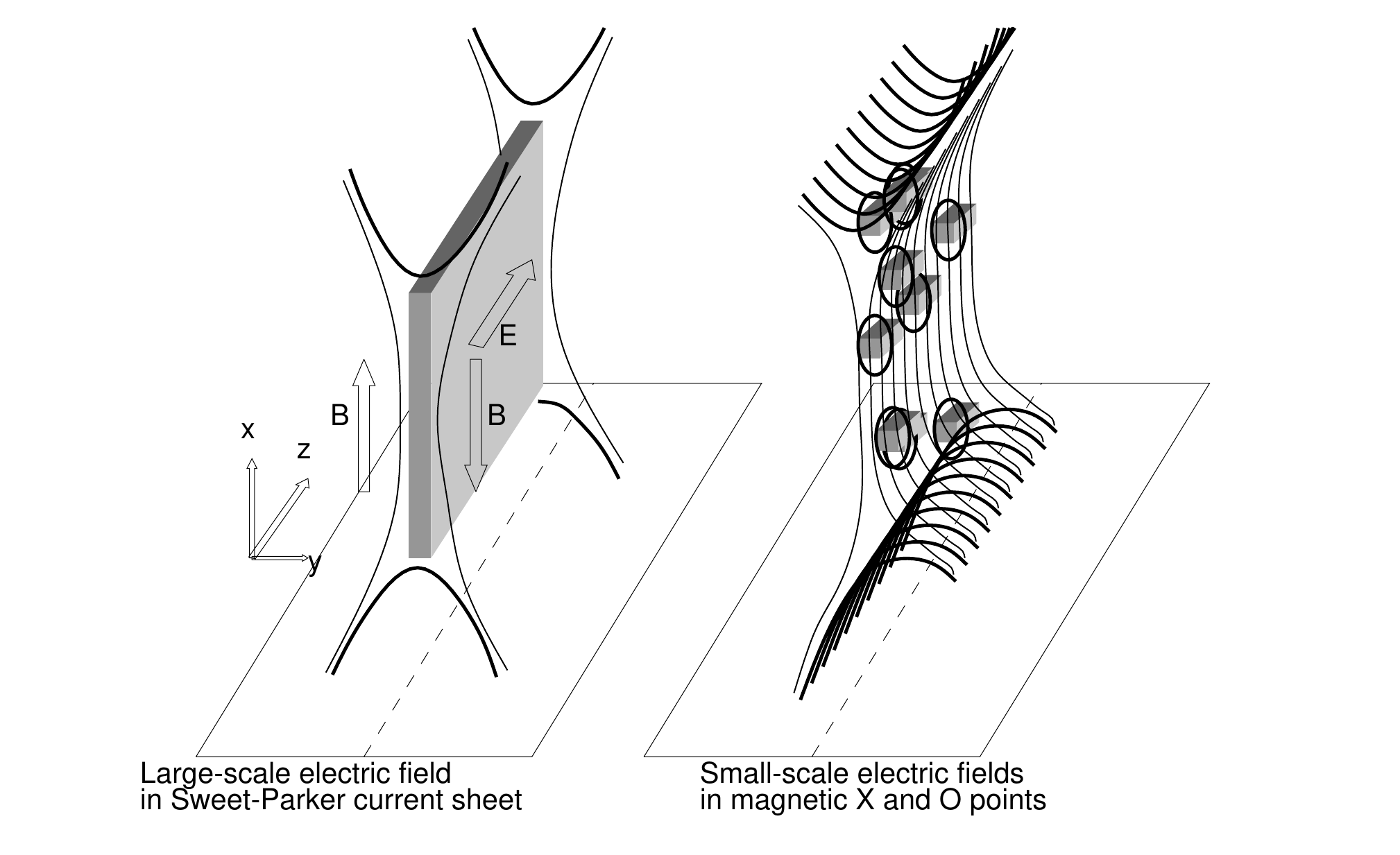}}
\caption{Paradigm shift of current sheet structure and 
associated particle acceleration regions: {\sl Left:} 
classical models assume large-scale electric fields based 
on Sweet$-$Parker magnetic reconnection, which have a much
larger extent in the x and z-direction than their width in 
the y-direction. {\sl Right:} theory and MHD simulations, 
however, imply small-scale electric fields in magnetic X-points
and coalescing islands with magnetic O-points (Aschwanden 2002).}
\end{figure}

\begin{figure}
\centerline{\includegraphics[width=1.0\textwidth]{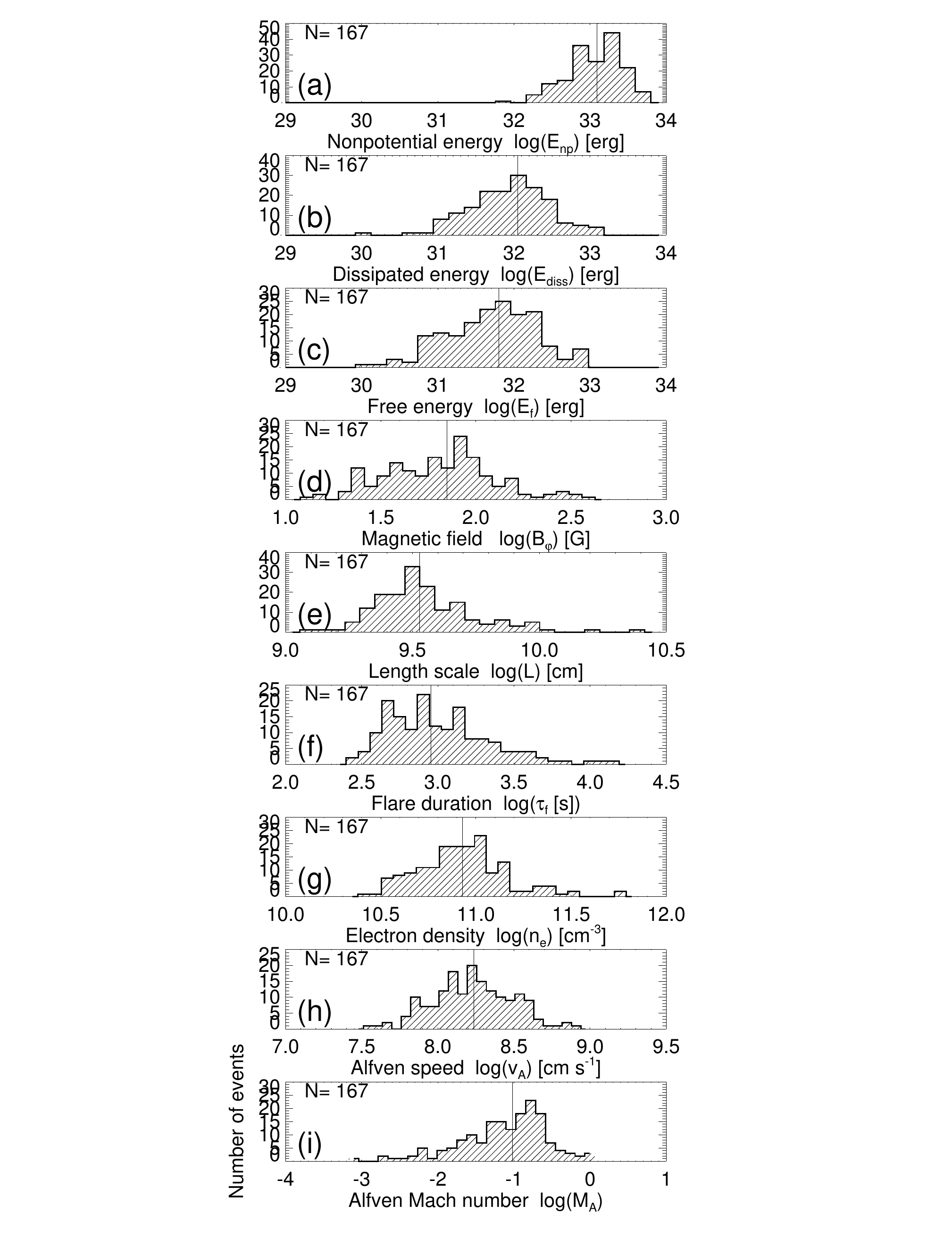}}
\caption{Histogrammed distributions of observables and derived
physical parameters of 162 analyzed M- and X-class flare events.
The median values are marked with a vertical line, see also Table 1.}
\end{figure}

\begin{figure}
\centerline{\includegraphics[width=1.0\textwidth]{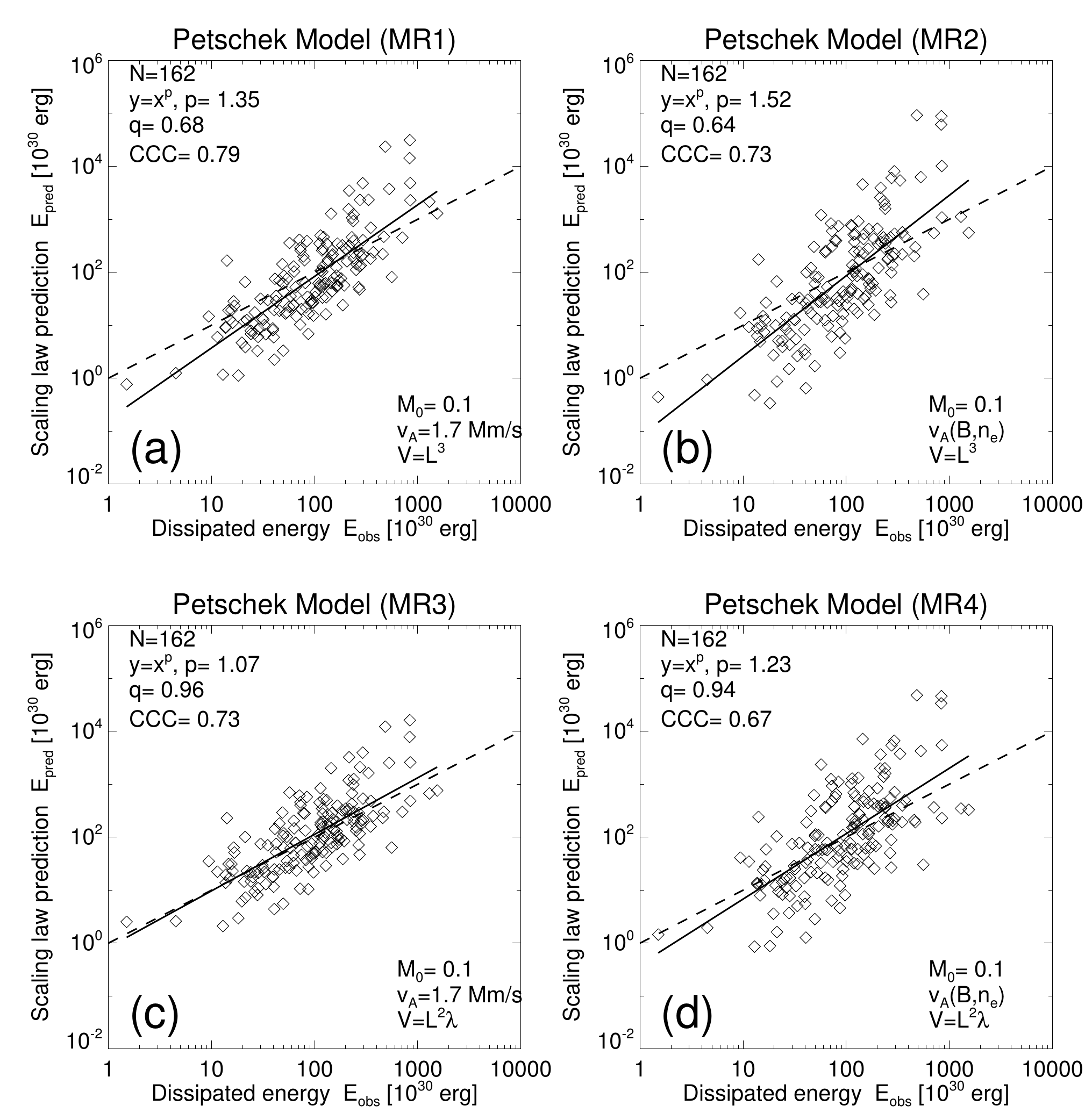}}
\caption{The scaling law prediction of four versions 
(MR1, MR2, MR3, MR4) of the Petschek model  
as a function of the observed dissipated 
energy $E_{diss}$ from a set of $N=162$ M- and X-class 
flare events (diamond symbols). The four model versions 
embody different scalings of the Alfv\'en velocity
($v_A=1700$ km s$^{-1}$ versus $v_A(B,n_e)$, or
volume scaling ($V=L^3$ versus $V^2 \lambda$),
and Mach number is $M_A = M_0 = 0.1$
The solid lines show linear regression fits, and the 
dashed lines represent equivalence between the observed 
dissipated energies and the predicted scaling law energies.}
\end{figure}

\begin{figure}
\centerline{\includegraphics[width=1.0\textwidth]{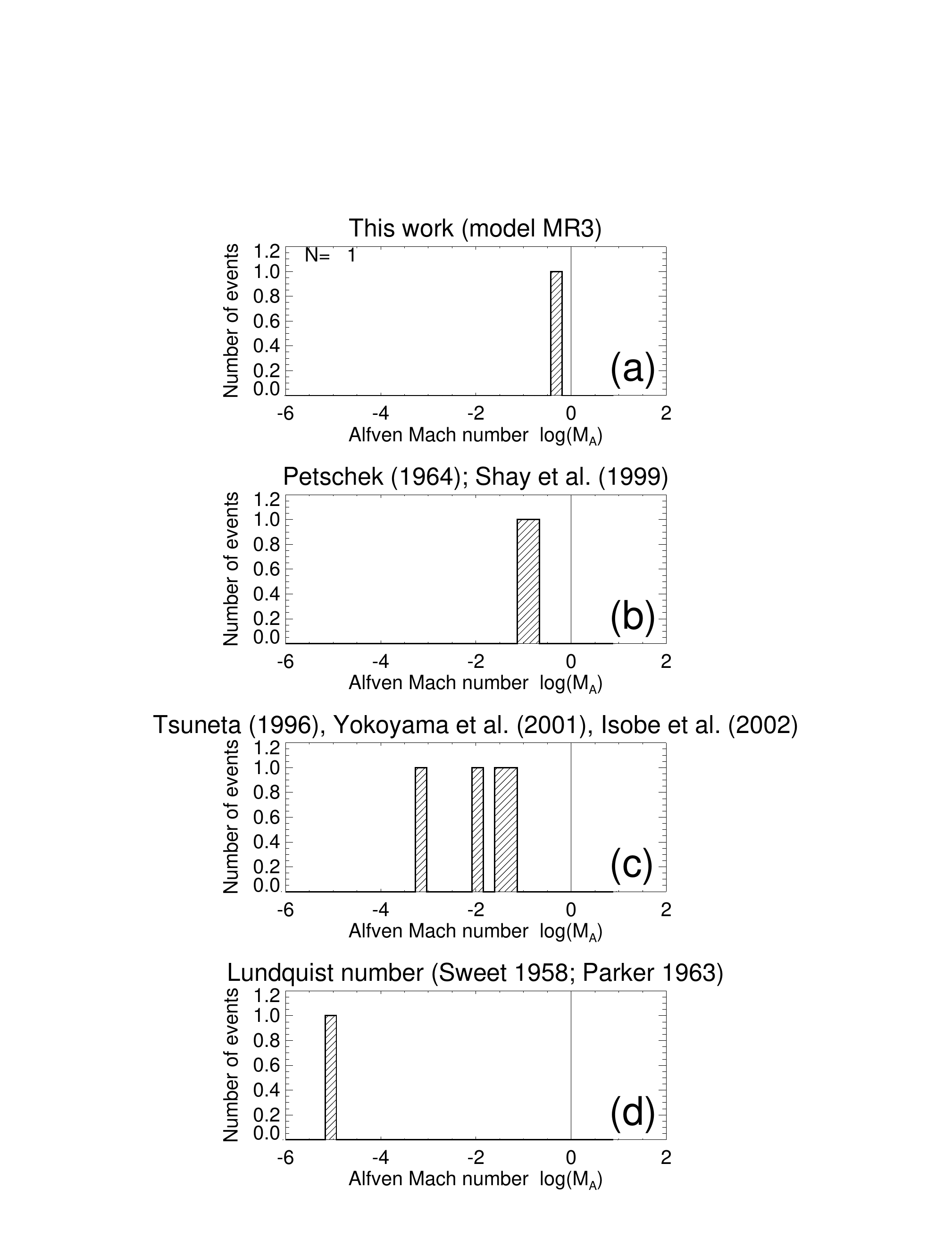}}
\caption{Theoretical (panels b and d) and observational 
estimates of the Alfv\'en Mach number $M_A$, derived from
Yohkoh/SXT observations (c) and from model MR3 in this work (a).}
\end{figure}

\end{document}